\documentclass{PoS}

\newcommand{\iep}{i\varepsilon}

\title{
\vspace{-3.5cm}
{\normalsize \rm $\phantom{.}$ \hfill Alberta Thy 01-10}
\vspace{3cm}\\
Two-loop corrections to the Lamb shift}

\ShortTitle{Two-loop corrections to the Lamb shift}

\author{Matthew Dowling, Jorge Mond{\'e}jar, \speaker{Jan H. Piclum} and
  Andrzej Czarnecki\\
  University of Alberta, Department of Physics, Edmonton, Alberta,
  Canada T6G 2G7\\
  E-mail: \email{mdowling@phys.ualberta.ca},
  \email{jmonde@phys.ualberta.ca}, \email{jpiclum@phys.ualberta.ca},
  \email{andrzej.czarnecki@ualberta.ca}}

\abstract{We present a new calculation of the order $\alpha^2 (Z
  \alpha)^5$ correction to the Lamb shift, which uses methods developed
  for multi-loop calculations.}

\FullConference{RADCOR 2009 - 9th International Symposium on Radiative Corrections
(Applications of Quantum Field Theory to Phenomenology) \\
                 October 25-30 2009\\
                 Ascona, Switzerland}
\begin{document}
\section{Introduction}

The Lamb shift in hydrogen was discovered in
1947~\cite{Lamb:1947}. Since then, developments in spectroscopy have led
to very precise experimental values for the $1S$ Lamb
shift~\cite{Berkeland:1995,Weitz:1995zz,Bourzeix:1996zz,Udem:1997zz,Schwob:1999zz},
making it the best test of Quantum Electrodynamics for an atom. On the
theory side, much effort has been put into the calculation of higher
order corrections to match the experimental precision
(cf. Ref.~\cite{Eides:2000xc} for a review of the theory of light
hydrogen-like atoms).

In the perturbative calculation it is important to correctly account for
the presence of three different small parameters. $Z\alpha$ describes
the binding effects of an electron to a nucleus with charge number
$Z$. Self-interactions of the electron lead to additional powers of
$\alpha$, but not of $Z$. Therefore, it is useful to keep a generic $Z$
to distinguish the two different effects, even though our main
application is hydrogen where $Z$ is equal to one. The third small
parameter is the ratio of electron and nucleus masses, $m/M$. The Lamb
shift is of order $\alpha(Z\alpha)^4$. Corrections also include
logarithms of $Z\alpha$ and $m/M$. At present, all second-order
corrections in $\alpha$ and $Z\alpha$ are known, as well as some third
order ones~\cite{Pachucki:2001zz,Eides:2006hg}.

Another important correction is due to the spatial distribution of the
nuclear charge. At the moment, the theoretical prediction for hydrogen is
limited by the uncertainty in the measurement of the proton root mean
square charge radius. However, new
measurements~\cite{Antognini:2005,Lauss:2009ng} of the Lamb shift in
muonic hydrogen are expected to improve the knowledge of this parameter,
soon. The advantage of muonic hydrogen is the greater sensitivity to the
proton charge distribution due to the larger mass, and thus stronger
binding, of the muon.

Here we present a new calculation of the second-order non-recoil
corrections of order $\alpha^{2}(Z\alpha)^{5}$. The full result for
these corrections was presented first in Ref.~\cite{Pachucki:1994} and
improved in Refs.~\cite{Eides:1995ey,Eides:1995gy}. Our
result~\cite{Dowling:2009md} is compatible with the previous ones and
has better precision.

\section{Calculation\label{sec::calc}}

\begin{figure}[b]
  \center
  \includegraphics[width=0.6\textwidth]{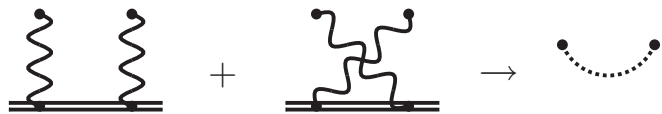}
  \caption{\label{fig::eff}The effective propagator for the interaction
    with the nucleus. Wavy and double lines denote photons and the
    nucleus, respectively.}
\end{figure}

\begin{figure}[t]
  \center
  \includegraphics[width=\textwidth]{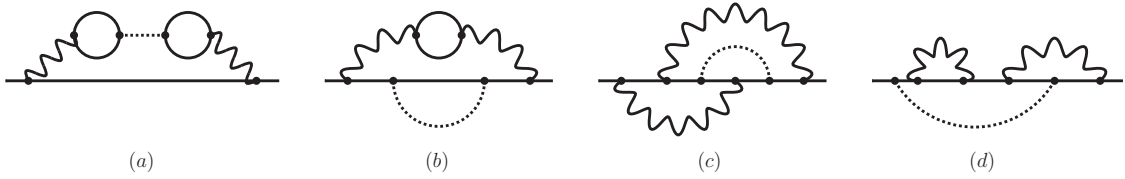}
  \caption{\label{fig::dias}Sample diagrams. Solid, wavy, and dotted
    lines denote electrons, photons, and the effective propagator,
    respectively.}
\end{figure}

Since we are only interested in non-recoil corrections, we consider the
scattering of an on-shell electron with an on-shell nucleus, where both
external momenta have vanishing space-like components. Furthermore, we
only have to calculate the leading term in the expansion around
$M\to\infty$. In this particular case we can simplify the interaction
with the nucleus when we consider the diagram with direct photon
exchange together with the diagram with crossed photon exchange, as
depicted in Fig.~\ref{fig::eff}. In the infinite mass limit, the
nucleus propagators become static propagators. However, due to the
different momentum flow in the two diagrams, the sum of the two becomes
the difference of two terms which differ only in the sign of the
$\iep$ prescription of these propagators. Neglecting the Dirac
structure, we have
\begin{equation}
  \frac{1}{(N+k)^2 - M^2 + \iep} + \frac{1}{(N-k)^2 - M^2 + \iep}
  \stackrel{M\to\infty}{\to} \frac{1}{2N\cdot k + \iep} -
  \frac{1}{2N\cdot k - \iep} = \frac{i\pi}{M} \delta(k^0) \,,
\end{equation}
where $N = (M,\vec{0})$ and $k$ are the nucleus and photon momentum,
respectively. The latter is considered to scale like the electron
mass, which is much smaller than the nucleus mass. Thus, we have to
compute three-loop electron self-energy diagrams with two photons and
one effective propagator, which is defined by Fig~\ref{fig::eff}. Sample
diagrams are depicted in Fig.~\ref{fig::dias}.

Our calculation proceeds as follows. We use {\tt QGRAF}~\cite{Nogueira:1991ex}
to generate the Feynman diagrams, and {\tt q2e} and
{\tt exp}~\cite{Harlander:1997zb,Seidensticker:1999bb} to turn them 
into {\tt FORM}~\cite{Vermaseren:2000nd} readable code. Finally, we use
{\tt MATAD3}~\cite{Steinhauser:2000ry} to do the Dirac algebra and
express all diagrams in terms of scalar integrals, using self-made
routines. The next step is the reduction to so-called master integrals
using integration-by-parts
identities~\cite{Tkachov:1981wb,Chetyrkin:1981qh}. For this we use the
program {\tt FIRE}~\cite{Smirnov:2008iw}, which is a {\tt Mathematica}
implementation of the so-called Laporta
algorithm~\cite{Laporta:1996mq,Laporta:2001dd}.

\begin{figure}[h]
  \center
  \includegraphics[width=0.8\textwidth]{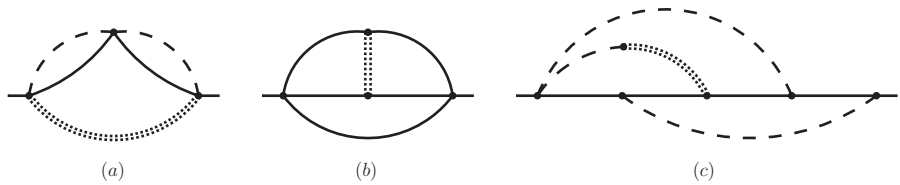}
  \caption{\label{fig::masters}Sample master integrals. Solid and dashed
    lines denote scalar massive and massless propagators,
    respectively. Dotted double lines denote the delta function.}
\end{figure}

Sample master integrals are depicted in Fig.~\ref{fig::masters}. We used
different methods for their evaluation. Simpler integrals like the one
in Fig.~\ref{fig::masters}$(a)$ were computed with the Mellin-Barnes
method~\cite{Smirnov:1999gc,Tausk:1999vh}, using the {\tt Mathematica}
packages {\tt MB}~\cite{Czakon:2005rk} and {\tt MBresolve}~\cite{Smirnov:2009up}.
For more complicated integrals, we used sector
decomposition~\cite{Binoth:2000ps,Binoth:2003ak}, as implemented in the
program {\tt FIESTA}~\cite{Smirnov:2008py,Smirnov:2009pb}. The most
complicated master integral is the one in
Fig.~\ref{fig::masters}$(c)$. Fortunately, this integral is finite, and
we were able to derive a Feynman-parameter representation which could be
integrated with the help of the {\tt Cuba}
library~\cite{Hahn:2004fe}. Results for all master integrals are given
in Ref.~\cite{Dowling:2009md}.

\section{Results\label{sec::res}}

In order to present our final result, we split the Feynman diagrams into
two subsets: diagrams with closed electron loops
(cf. Fig.~\ref{fig::dias}$(a)$ and $(b)$) are denoted by the subscript
$vp$, and diagrams without closed electron loops
(cf. Fig.~\ref{fig::dias}$(c)$ and $(d)$) are denoted by the subscript
$nvp$. Our results for the corresponding energy shifts are
\begin{eqnarray}
  \delta E_{vp} & = & \frac{\alpha^{2}(Z\alpha)^{5}}{\pi n^{3}} \left(
  \frac{\mu}{m} \right)^{3} m \, [ 0.86281422(3) ] \,, \\
  \delta E_{nvp} & = & \frac{\alpha^{2}(Z\alpha)^{5}}{\pi n^{3}} \left(
  \frac{\mu}{m} \right)^{3} m \, [ -7.72381(4) ] \,,
\end{eqnarray}
where $\mu = mM/(m+M)$ denotes the reduced mass of the atom, and $n$ is
the principal quantum number. Results for individual diagrams can be
found in Ref.~\cite{Dowling:2009md}.

The best results for the two subsets prior to our calculation have been
published in Ref.~\cite{Pachucki:1993zz} (cf. Ref.~\cite{Eides:2000xc}
for references of partial results) and Ref.~\cite{Eides:1995ey},
respectively. Our results agree with the previous ones within the error
bars. However, we improve the precision by two orders of magnitude for
$\delta E_{vp}$, and a little over one order of magnitude for $\delta
E_{nvp}$.

The total result reads
\begin{equation}
  \delta E = \frac{\alpha^{2}(Z\alpha)^{5}}{\pi n^{3}} \left(
  \frac{\mu}{m} \right)^{3} m \, [ -6.86100(4) ] \,,
\end{equation} 
and the corresponding energy shifts for the $1S$ and the $2S$ states in
hydrogen are
\begin{eqnarray}
  \delta E_{1S} & = & -296.866(2)\,\mbox{kHz}\,,\\
  \delta E_{2S} & = & -37.1082(3)\,\mbox{kHz}\,.
\end{eqnarray}

\acknowledgments
We are grateful to A.V. and V.A. Smirnov for providing us with a beta
version of {\tt FIESTA 2} prior to publication. JHP thanks the
organisers of ``RADCOR 2009'' for an interesting conference. This work
was supported by the Natural Sciences and Engineering Research Council
of Canada and the Alberta Ingenuity Foundation. The Feynman diagrams
were drawn using \texttt{Axodraw}~\cite{Vermaseren:1994je} and
\texttt{Jaxodraw 2}~\cite{Binosi:2008ig}.

\end{document}